\documentclass[12pt]{iopart}

\usepackage{graphicx}
\usepackage{amsthm}
\usepackage{amsfonts}
\usepackage{amssymb}
\usepackage{graphicx}

\usepackage{indentfirst}

\usepackage{url}

\begin{document}
\title[Circularly Polarized Molecular HHG]{Circularly Polarized Molecular High Harmonic Generation Using a Bicircular Laser}

\author{F. Mauger$^{1,2}$, A.~D. Bandrauk$^{1}$ and T. Uzer$^{3}$}
\address{$^{1}$Laboratoire de Chimie Th\'eorique, Facult\'e des Sciences, Universit\'e de Sherbrooke, Sherbrooke, Qu\'ebec, Canada J1K 2R1, \\
$^{2}$Department of Physics and Astronomy, Louisiana State University, Baton Rouge, Louisiana 70803-4001, USA,\\
$^{3}$School of Physics, Georgia Institute of Technology, Atlanta, GA 30332-0430, USA}
\ead{fmauger@phys.lsu.edu}


\begin{abstract}
We investigate the process of circularly polarized high harmonic generation in molecules using a bicircular laser field. In this context, we show that molecules offer a very robust framework for the production of circularly polarized harmonics, provided their symmetry is compatible with that of the laser field. Using a discrete time-dependent symmetry analysis, we show how all the features (harmonic order and polarization) of spectra can be explained and  predicted. The symmetry analysis is generic and can easily be applied to other target and/or field configurations.
\end{abstract}


\submitto{\JPB}
\maketitle


In the past few decades, strong-field physics which studies the interaction between strong and ultra-short laser pulses and atoms and molecules has drawn an increasing level of interest for the perspectives it offers to probe the organization of matter~\cite{Beck12,Vrak14,Itat04,Haes10,Shao10}, at the scale of the electronic dynamics in this systems, thus giving birth to attosecond science~\cite{Krau09}. An efficient means of obtaining such ultra-fast pulses relies on the production and manipulation of high-harmonic generation (HHG) emission. For atomic targets, our understanding of HHG relies on the conversion of a large number of laser photons, previously absorbed by an ionized electron, into a single harmonic photon upon recollision~\cite{Cork93,Scha93,Kuch87}. Beyond atoms, similar HHG have been reported and studied for molecular targets where one can take advantage of the more complex geometry~\cite{Zhou09,Yuan12} compared with the atomic counterpart.

\begin{figure}
	\centering
		\includegraphics[width=.45\linewidth]{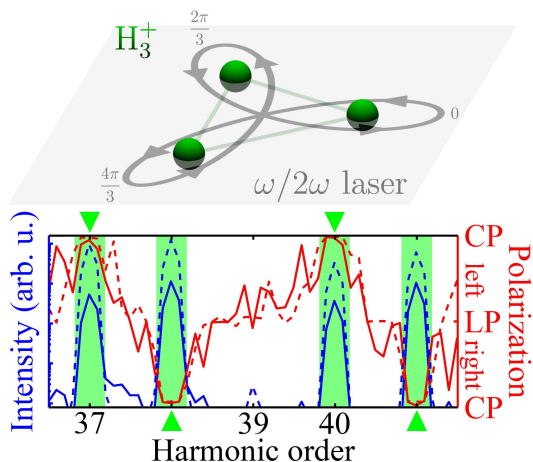}
	\caption{\label{fig:H3+}
	Upper part: Equilateral ${\rm H}_{3}^{+}$ molecular ion (dark green balls) and counter-rotating bicircular $\omega/2\omega$ laser field (light gray curves -- the arrows indicate the temporal evolution of the field).
	Lower part: Associated HHG spectrum (dark blue curves, left hand axis) and polarization (light red curves, right hand axis) for $2$ dimensional (solid curves) and $3$ dimensional (dashed curves) models~\cite{Potential_note,Laser_note}. The predicted harmonic orders and polarization (from the symmetry analysis) are labeled with the light green stripes and upper/lower triangles respectively.
	As illustrated in the upper part, the molecule and laser have compatible symmetries such that, each $1/3$ period of the laser, the emission signal is repeated with $2\pi/3$ rotation. The combination of all those emissions leads to specific circularly polarized harmonics, as shown in the lower part.}
\end{figure}

Numerical investigations of molecular circularly polarized (CP) HHG have been reported for specific geometrical (stretched molecules) and/or laser field (combination of elliptic/CP and static laser fields, carrier envelope phase~\ldots) configurations~\cite{Yuan12}. Recently, the interest in CP harmonics have been revived by experimental work using an atomic target and a counter rotating bicircular laser~\cite{Flei14}, opening the way for dichroism studies. Using electric dipole selection rules (EDSR), the properties of such bicircular atomic spectra have been explained in a photon picture~\cite{Eich95,Milo00,Pisa14}. Here we investigate the process of molecular HHG with bicircular laser fields. For molecular systems the stakes are high given that EDSR do not apply (because of the absence of rotational symmetry) while the more complex geometry offers perspectives for richer features compared to atoms~\cite{Zhou09,Yuan12}. We use discrete time-dependent symmetry (DTDS)~\cite{Cecc01} analysis and show that molecular CP HHG can be obtained in generic configurations of symmetric molecules, as shown in figure~\ref{fig:H3+}.

From the theoretical and numerical points of view, most HHG models disregard the photon quantum nature of light and compute spectra from the dipole (acceleration) of the electronic wavefunction~\cite{Haes11}
\begin{equation} \label{eq:HHG_spectrum}
	{\bf R}_{\rm HHG} = 
		\mathcal{F}\left[\ddot{\bf d}\left(t\right)\right]\left(\nu\right) = 
		\mathcal{F}\left[
			\left\langle\psi\left(t\right)\left|\hat{\bf a}\left(t\right)\right|\psi\left(t\right)\right\rangle
		\right]\left(\nu\right),
\end{equation}
where $\mathcal{F}$ denotes the Fourier transform and we use the bracket notation. Here $\left|\psi\left(t\right)\right\rangle$ is the numerical solution of the time-dependent Schr\"{o}dinger equation (TDSE) for an isolated single active electron (SAE) model taken in the dipole approximation. In the length gauge and using atomic units (unless otherwise specified), the corresponding Hamiltonian operator reads
\begin{equation} \label{eq:Hamiltonian}
	\hat{\mathcal{H}}\left({\bf x},t\right) =  
		\underbrace{-\frac{\Delta}{2} + \mathcal{V}\left({\bf x}\right)}_{\hat{\mathcal{H}}_{0}\left({\bf x}\right)} 
		+ {\bf E}\left(t\right)\cdot\hat{\bf x},
\end{equation}
where $\hat{\mathcal{H}}_{0}$ is the free-field molecular Hamiltonian, $\mathcal{V}$ is the effective (SAE) potential~\cite{Potential_note} and ${\bf E}$ is the laser electric field. 

We present the DTDS analysis for a coplanar bicircular laser field with respective frequencies $\omega_{1}$ and $\omega_{2}$, which has drawn experimental and theoretical interest in the past~\cite{Eich95,Milo00,Cecc01} but it can easily be generalized to other laser configurations. We take the counter-rotating case as a reference for the laser such that, in the $\left(x,y\right)$ polarization plane, the bicircular electric field reads 
\begin{equation} \label{eq:Laser_field}
	{\bf E}\left(t\right) = E_{0} f\left(t\right) \left( \begin{array}{c}
			\cos \omega_{1} t + \cos \omega_{2} t \\
			\sin \omega_{1} t - \sin \omega_{2} t
		\end{array}\right),
\end{equation}
where $E_{0}$ and $f$ are the (common) peak field amplitude and envelope respectively\footnote{Absolute phases for each color component have been canceled with an appropriate time shift and frame rotation.}. In most cases, the beating between the two color components induces a periodic vanishing of the laser field (see the light gray curve in the upper part of figure~\ref{fig:H3+}) which is naturally associated with recollision and HHG. This becomes obvious in a rotating frame at frequency $\Delta\omega=\left(\omega_{1}-\omega_{2}\right)/2$, which maps Hamiltonian~(\ref{eq:Hamiltonian}) to~\cite{Zuo95}
\begin{equation} \label{eq:Rotating_frame_Hamiltonian}
	\hat{\tilde{\mathcal{H}}}\left(\tilde{\bf x},t\right) = 
		\hat{\tilde{\mathcal{H}}}_{0}\left(\tilde{\bf x},t\right) - 
		\Delta\omega \hat{\tilde{\mathcal{L}}}_{z} +
		2 E_{0}f\left(t\right)\cos \overline{\omega} t \hat{\tilde{x}},
\end{equation}
where tildes stand for rotating frame coordinates, $\hat{\tilde{\mathcal{L}}}_{z}=-i\left(\hat{\tilde{x}}\partial_{\tilde{y}}-\hat{\tilde{y}}\partial_{\tilde{x}}\right)$ is the angular momentum operator and $\overline{\omega}=\left(\omega_{1}+\omega_{2}\right)/2$. From the lower part of figure~\ref{fig:H3+} we notice that only some harmonics, with a well defined polarization, are present in the spectrum as we explain in what follows using DTDS analysis.

The role of dynamical symmetries in HHG computations has been identified previously considering eigenstate decomposition of the wavefunction using free-field~\cite{Nils02} or Floquet~\cite{Cecc01,Alon98} state. Here we present a fully \emph{non-perturbative} analysis of HHG and begin with the laser component of Hamiltonian~(\ref{eq:Hamiltonian}). For the theoretical analysis, and without loss of generality, we assume $0<\left|\omega_{1}\right|\leq\left|\omega_{2}\right|$ in equation~(\ref{eq:Laser_field}). Finally, in order to impose some periodicity on the system, we assume the two color components locked, i.e., $\omega_{2}/\omega_{1}=k_{2}/k_{1}$ with $k_{1}$ and $k_{2}$ nonzero integers ($k_{2}>0$ thus $k_{1}+k_{2}\geq0$) and co-prime (irreducible fractional form), which is not a strong hypothesis since rational numbers are dense in real numbers $\mathbb{R}$. We illustrate the laser field in counter-rotating $\omega/2\omega$ ($k_{2}=2k_{1}$) with a light gray curve in the upper part of figure~\ref{fig:H3+}. Putting aside the envelope ($f=1$ although numerical simulations of the TDSE fully account for it), basic trigonometric algebra shows that the electric field~(\ref{eq:Laser_field}) is $2\pi k_{1}/\omega_{1}$-time periodic and exhibits a $k_{1}+k_{2}$-fold rotational symmetry (for $k_{1}+k_{2}\neq0$)
\begin{equation} \label{eq:Electric_field_symmetry}
	{\bf E}\left(t+n\frac{\theta}{\omega_{1}}\right) = {\bf \Omega}\left(n\theta\right){\bf E}\left(t\right), \ \ \ 
		\theta=2\pi\frac{k_{1}}{k_{1}+k_{2}},
\end{equation}
where ${\bf \Omega}\left(\alpha\right)$ is the rotation matrix around the $z$ axis with angle $\alpha$, for all integer $n$. For example, in the counter-rotating $\omega/2\omega$ configuration, this leads to an overall $3$-fold symmetry (again, see the light gray curve in the upper part of figure~\ref{fig:H3+}).

Moving to the molecular part of Hamiltonian~(\ref{eq:Hamiltonian}) we assume that the target also presents some degree of symmetry which is compatible with the field own symmetry. More specifically, we assume that the potential SAE (effective) potential presents a $k$-fold symmetry
$
	\hat{\mathcal{H}}_{0}\left({\bf x}\right) = \hat{\mathcal{H}}_{0}\left({\bf \Omega}\left(-2\pi/k\right){\bf x}\right).
$
Then, the overall system ``molecule+laser field'' presents a $\tilde{k}$-fold symmetry, where $\tilde{k}$ as the greatest common divisor between $k$ and $k_{1}+k_{2}$
\begin{equation} \label{eq:Hamiltonian_symmetry}
	\hat{\mathcal{H}}\left({\bf x},t+n\frac{\theta}{\omega_{1}}\right) = 
		\hat{\mathcal{H}}\left({\bf \Omega}\left(-n\theta\right){\bf x},t\right), \ \ \ 
		\tilde{\theta}=2\pi\frac{k_{1}}{\tilde{k}},
\end{equation}
as deduced from~(\ref{eq:Electric_field_symmetry}), for all integer $n$. As can be seen in the upper part of figure~\ref{fig:H3+}, the combination of the equilateral ${\rm H}_{3}^{+}$ molecular ion~\cite{Potential_note} and counter-rotating $\omega/2\omega$ laser field is an example of such a compatible symmetry. Experimentally and numerically, HHG corresponds to a range of intensities where ionization is moderate as HHG vanishes for depletion of the ground state. Therefore it is reasonable to assume that the wavefunction repeats itself over time, following a similar symmetry to equation~(\ref{eq:Hamiltonian_symmetry}) -- up to a phase that disappears in the dipole (acceleration) computation -- and in turn, so does the dipole (acceleration). We have checked numerically~\cite{Potential_note} the limiting factor of ionization by observing that the properties of HHG spectra described in what follows degrade when entering a (strong) ionization regime, due to the loss of periodicity associated with ground state depletion.

We combine the symmetry condition~(\ref{eq:Hamiltonian_symmetry}) with the definition of the Fourier transform~(\ref{eq:HHG_spectrum}) over one period of the laser field
\begin{eqnarray*}
	{\bf R}_{\rm HHG} \left(\nu\right) & = & 
		\frac{\omega_{1}}{2\pi k_{1}}\int_{0}^{2\pi\frac{k_{1}}{\omega_{1}}}{\left(\begin{array}{c}
			\ddot{d}_{x}\left(t\right) \\ \ddot{d}_{y}\left(t\right)
		\end{array}\right){\rm e}^{-i\nu t}\ dt}, \\ & = &
		\frac{\omega_{1}}{2\pi k_{1}}\sum_{k=0}^{\tilde{k}-1}{
			\int_{k\frac{\tilde{\theta}}{\omega_{1}}}^{\left(k+1\right)\frac{\tilde{\theta}}{\omega_{1}}}{\left(\begin{array}{c}
				\ddot{d}_{x}\left(t\right) \\ \ddot{d}_{y}\left(t\right)
			\end{array}\right) {\rm e}^{-i\nu t}\ dt}
		}, \\ & = & 
		\frac{\omega_{1}}{2\pi k_{1}}\sum_{k=0}^{\tilde{k}-1}{{\rm e}^{-i\nu\frac{\tilde{\theta}}{\omega_{1}}}
			\int_{0}^{\frac{\tilde{\theta}}{\omega_{1}}}{\left(\begin{array}{c}
				\ddot{d}_{x}\left(t+k\frac{\tilde{\theta}}{\omega_{1}}\right) \\
				\ddot{d}_{y}\left(t+k\frac{\tilde{\theta}}{\omega_{1}}\right)
			\end{array}\right){\rm e}^{-i\nu t}\ dt}
		},
\end{eqnarray*}
where $d_{x}$ and $d_{y}$ are the dipole components in the polarization plane and we have used simple time translation in each component of the sum in the final equality. Then, using the symmetry condition~(\ref{eq:Hamiltonian_symmetry}) for the dipole (acceleration) we factorize the spectrum
$$
	{\bf R}_{\rm HHG} = \frac{1}{\tilde{k}}\sum_{k=0}^{\tilde{k}-1}{
			{\rm e}^{-i\nu\frac{\tilde{\theta}}{\omega_{1}}}\left(\begin{array}{rr}
				\cos k \tilde{\theta} & -\sin k \tilde{\theta} \\ \sin k \tilde{\theta} & \cos k \tilde{\theta}
			\end{array}\right)
		}\underbrace{
			\frac{\omega_{1}}{\tilde{\theta}}\int_{0}^{\frac{\tilde{\theta}}{\omega_{1}}}{\left(\begin{array}{c}
				\ddot{d}_{x}\left(t\right) \\ \ddot{d}_{y}\left(t\right)
			\end{array}\right){\rm e}^{-i\nu t}\ dt}
		}_{
			R_{x}^{0} {\bf e}_{x} + R_{y}^{0} {\bf e}_{y}
		},
$$
where ${\bf e}_{x}$ and ${\bf e}_{y}$ are the units vectors in the $x$ and $y$-directions respectively and we notice that the (constant) factors $R_{x}^{0}$ and $R_{y}^{0}$ have been factorized out of the previous sum. Then, using the exponential definition of the cosine and sine function, and after factorization, the HHG spectrum summarizes to
\begin{equation} \label{eq:HHG_spectrum_factorization}
	{\bf R}_{\rm HHG} = \frac{1}{\tilde{k}}\sum_{k=0}^{\tilde{k}-1}\left(\begin{array}{c}
			\frac{R_{x}^{0}+i R_{y}^{0}}{2}  {\rm e}^{ ik\tilde{\theta}\left(1-\frac{\nu}{\omega_{1}}\right)} +
			\frac{R_{x}^{0}-i R_{y}^{0}}{2}  {\rm e}^{-ik\tilde{\theta}\left(1+\frac{\nu}{\omega_{1}}\right)} \\
			\frac{R_{x}^{0}+i R_{y}^{0}}{2i} {\rm e}^{ ik\tilde{\theta}\left(1-\frac{\nu}{\omega_{1}}\right)} -
			\frac{R_{x}^{0}-i R_{y}^{0}}{2i} {\rm e}^{-ik\tilde{\theta}\left(1+\frac{\nu}{\omega_{1}}\right)} 
		\end{array}\right),
\end{equation}
and the geometric sums can be computed analytically leading to
\begin{equation} \label{eq:HHG_spectrum_sums}
	\sum_{k=0}^{\tilde{k}-1}{{\rm e}^{ik\tilde{\theta}\left(1\pm\frac{\nu}{\omega_{1}}\right)}} = \left\{\begin{array}{l}
			\frac{
				1-{\rm e}^{i2\pi k_{1} \left(1\pm\frac{\nu}{\omega_{1}}\right)}
			}{
				1-{\rm e}^{i\tilde{\theta} \left(1\pm\frac{\nu}{\omega_{1}}\right)}
			} \ \ \ {\rm for} \ \ \ \tilde{\theta}\left(1\pm\frac{\nu}{\omega_{1}}\right)\notin2\pi\mathbb{Z} \\
			\tilde{k} \ \ \ {\rm otherwise}
		\end{array}\right. ,
\end{equation}
where $\mathbb{Z}$ is the set of integers.

We begin the DTDS analysis with the generic symmetry case $\tilde{k}>2$, e.g., for the ${\rm H}_{3}^{+}$ molecular ion and counter-rotating $\omega/2\omega$ laser configuration as illustrated in the upper panel of figure~\ref{fig:H3+}. Since the two frequencies $\omega_{1}$ and $\omega_{2}$ are mode locked, one can define a common fundamental frequency $\overline{\omega}=\omega_{1}/k_{1}=\omega_{2}/k_{2}$ and from the overall ``molecule+laser field'' symmetry order we define the integer $\tilde{n}$ such that $k_{1}+k_{2}=\tilde{n}\tilde{k}$. Already, from the trigonometric sums~(\ref{eq:HHG_spectrum_sums}) we see that all the harmonics $\mathbb{Z}\overline{\omega}\backslash\tilde{k}\mathbb{Z}\overline{\omega}\pm\omega_{1}$ vanish from the spectrum. For the other ones, with $\tilde{k}\mathbb{Z}\overline{\omega}+\omega_{1}$ the spectrum~(\ref{eq:HHG_spectrum_factorization}) is such that $R_{x}/R_{y}=i$ which is characteristic of left (counterclockwise) CP emission wile with $\tilde{k}\mathbb{Z}\overline{\omega}-\omega_{1}$ we have $R_{x}/R_{y}=-i$ corresponding to right (clockwise) CP\footnote{We assume $\omega_{1}>0$, otherwise the respective helicities for the $\tilde{k}\mathbb{Z}\overline{\omega}\pm\omega_{1}$ have to the inverted.}; in other words, the HHG spectrum is circularly polarized with alternating helicities. Going back to our molecular model ($\tilde{k}=3$ and $\tilde{n}=1$) we predict that $\left(3n+1\right)\omega$ and $\left(3n-1\right)\omega$ are left and right CP respectively while $3n\omega$ harmonics are absent from the spectrum. Looking at the lower part of figure~\ref{fig:H3+}, we have a numerical confirmation of these predictions for two (continuous lines) and three (dash) dimensional simulations.

\begin{figure}
	\centering
		\includegraphics[width=.45\linewidth]{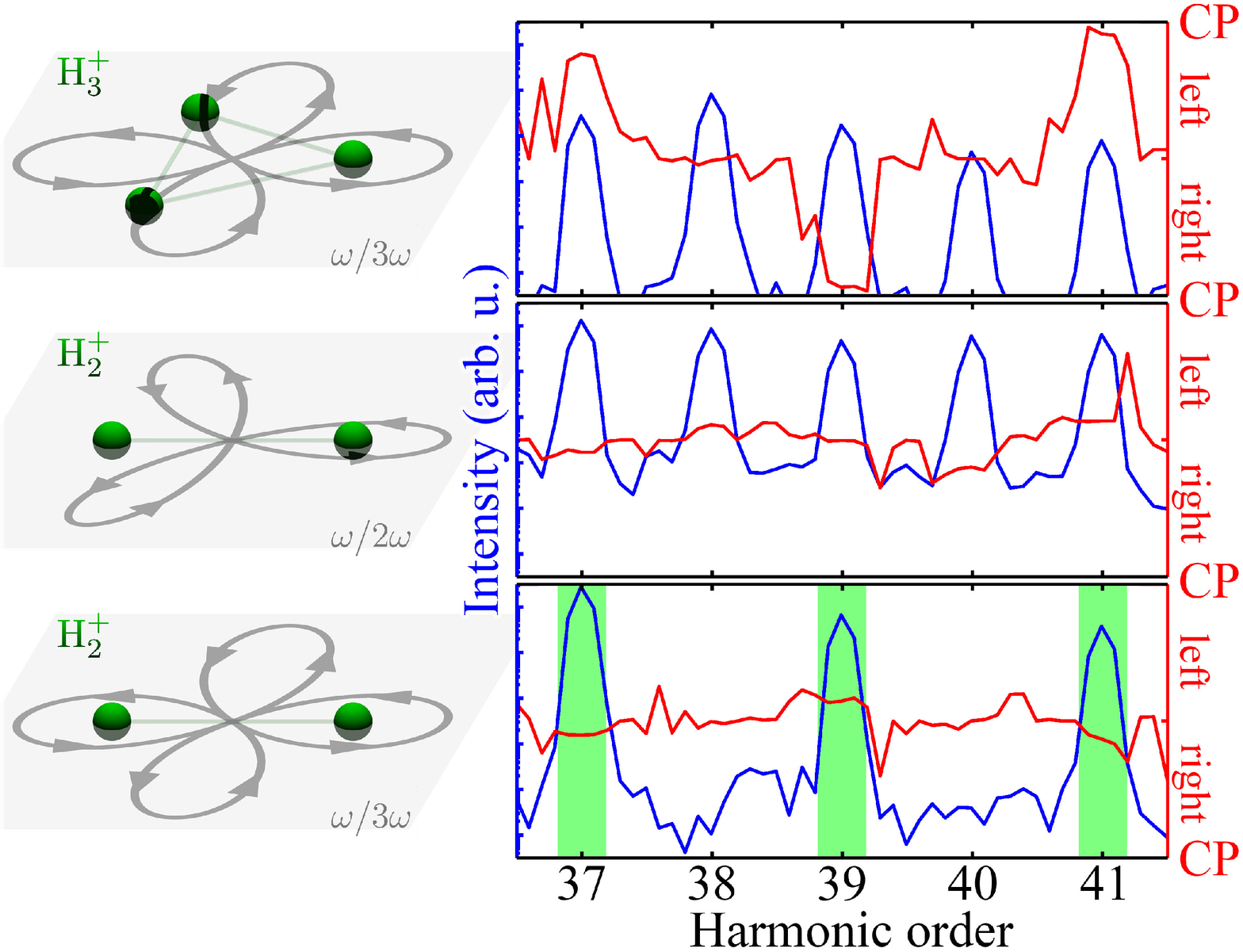}
	\caption{\label{fig:H2+}
	Same as figure~\ref{fig:H3+} for the ${\rm H}_{3}^{+}$ molecular ion with counter-rotating bicircular $\omega/3\omega$ laser field (upper part), ${\rm H}_{2}^{+}$ with $\omega/2\omega$ (middle) and ${\rm H}_{2}^{+}$ with $\omega/3\omega$ (lower)~\cite{Potential_note,Laser_note}.}
\end{figure}

The symmetry of the molecular system and laser field do not always match and may cancel each other, leading to $\tilde{k}=1$.  This is for instance the case for the ${\rm H}_{3}^{+}$ molecular ion with counter-rotating $\omega/3\omega$ laser field ($k=3$ and $k_{1}+k_{2}=4$ -- see upper part of figure~\ref{fig:H2+}) or ${\rm H}_{2}^{+}$ with $\omega/2\omega$ ($k=2$ and $k_{1}+k_{2}=3$ -- middle panel of the figure). With $\tilde{k}=1$ we see that the sums in equation~(\ref{eq:HHG_spectrum_sums}) are reduced to a single term and destructive/constructive interference effects are lost. As a consequence, all harmonics are made available to the HHG spectrum with arbitrary polarization. Here again, this is confirmed numerically in the upper right part of figure~\ref{fig:H2+} where all harmonics are present in the spectrum.

Finally, among all symmetry orders, the configuration $\tilde{k}=2$ offers an intermediate level of prediction. This minimum symmetry order allows constructive and destructive interferences in the sums~(\ref{eq:HHG_spectrum_sums}) leading to only odd harmonics of the common fundamental frequency $\left(2\mathbb{Z}+1\right)\overline{\omega}$ to be allowed in the spectrum. However, compared to the generic symmetry condition discussed previously, the two subsets $2\mathbb{Z}\overline{\omega}+\omega_{1}$ and $2\mathbb{Z}\overline{\omega}-\omega_{1}$ (respectively left and right CP) lead to the same harmonics in the overall spectrum which, in general, prevents drawing any conclusions on the polarization of those harmonics. With the ${\rm H}_{2}^{+}$ molecular ion model, the minimum symmetry configuration is obtained, e.g., with counter-rotating $\omega/3\omega$ CP fields, as shown in the lower left part of figure~\ref{fig:H2+}. Then looking at the associated HHG spectra (lower right part), we have the confirmation that only odd harmonics appear in the spectrum.
Note that time-dependent or rotational symmetries are not the only discrete symmetries one can use to draw predictions on the HHG spectrum. For instance, with a linearly polarized laser fields and linear molecules aligned parallel or orthogonal to the laser polarization, the system presents a symmetry by reflexion about the polarization axis. In turn, this leads to vanishing dipole moment in that specific direction resulting in linearly polarized harmonics (sharing the same polarization direction as the driving laser).

In conclusion, we have shown that molecular systems irradiated with a bicircular laser field offer a very robust framework for the generation of selected circularly polarized (CP) harmonics, provided their symmetry order is compatible with that of the laser. Using the \emph{non-perturbative} discrete time-dependent symmetry (DTDS) analysis, we have shown how these harmonics, and their polarization, can easily be predicted. Beyond the scope of the present manuscript, the DTDS is relatively generic and can easily be extended to other target and/or laser configurations. For example, atomic systems can be seen as a molecule with vanishing internuclear distance, i.e., the united atom limit. In this situation, the total symmetry of the ``atom+laser field'' always matches that of the laser alone because of the rotational symmetry of the atomic target. Then, applying a symmetry analysis almost identical to the one described previously we recover the predictions of the (perturbative) photon picture where only harmonics $\left(n\pm1\right)\omega_{1}+n\omega_{2}$ are allowed in the HHG spectrum~\cite{Eich95,Cecc01}, with opposite CP helicities, for $k_{1}+k_{2}>2$. Note though that in the case of a single CP field ($k_{1}+k_{2}=0$) the photon picture forbids the generation of harmonics altogether while the symmetry analysis imposes no restrictions leading to possibly all harmonics, as shown in~\cite{Maug14}.

%
\ack
F.M.\ and A.D.B\ thank S.~Chelkowski for enlightening discussions; F.M.\ thanks M.~Gaarde and K.J.~Schafer for enlightening discussions.
F.M.\ and A.D.B.\ thank RQCHP and Compute Canada for access to massively parallel computer clusters and the CIPI for financial support in its ultrafast science program.
F.M.\ and A.D.B.\ acknowledge financial support from the Centre de Recherches Math\'ematiques. F.M.\ acknowledges financial support from the Merit Scholarship Program for Foreign Student from the MESRS of Quebec. A.D.B.\ acknowledges financial support from the Canada Research Chair.
T.U.\ acknowledges funding from the NSF.
This research was supported in part by the National Science Foundation under Grant No. NSF PHY11-25915.



\begin{thebibliography}{99}
\bibitem{Laser_note}
For numerical simulations reported in this paper, we use a trapezoid envelope
  with $2$ fundamental laser period ramp up and $10$ cycle constant plateau. We
  restrict the HHG analysis to the plateau and use a Hanning
  window~\cite{MesPowSpec} to avoid spurious frequencies in the Fourier
  transform~(\ref{eq:HHG_spectrum}) due to the lack of periodicity of the
  dipole (acceleration) signal. The bicircular laser field is set with
  $780~{\rm nm}$ fundamental frequency and $3\times10^{14}~{\rm W}\cdot{\rm
  cm}^{-2}$ intensity. We have checked the robustness of results with the laser
  parameters and envelope.
\end{thebibliography}

\end{document}